\documentclass[12pt,preprint]{aastex}

\shorttitle{Spitzer observations of IRAS~16293-2422}
\shortauthors{J{\o}rgensen et al.}

\begin{document}

\title{Protostellar holes: Spitzer Space Telescope observations of the
  protostellar binary IRAS~16293-2422}

\author{Jes K. J{\o}rgensen\altaffilmark{1}, Fred
  Lahuis\altaffilmark{2,3}, Fredrik L. Sch\"{o}ier\altaffilmark{4},
  Ewine F. van Dishoeck\altaffilmark{2}, Geoffrey
  A. Blake\altaffilmark{5}, A.~C. Adwin Boogert\altaffilmark{6},
  Cornelis P. Dullemond\altaffilmark{7}, Neal J. Evans
  II\altaffilmark{8}, Jacqueline E. Kessler-Silacci\altaffilmark{8},
  and Klaus M. Pontoppidan\altaffilmark{2}}

\altaffiltext{1}{Harvard-Smithsonian Center for Astrophysics, 60 Garden Street, MS42, Cambridge, MA 02138 ({\tt jjorgensen@cfa.harvard.edu})}
\altaffiltext{2}{Leiden Observatory, PO Box 9513, 2300 RA Leiden, The Netherlands ({\tt freddy@sron.rug.nl; ewine, pontoppi@strw.leidenuniv.nl})}
\altaffiltext{3}{SRON National Institute for Space Research, PO Box 800, 9700 AV Groningen, The Netherlands}
\altaffiltext{4}{Stockholm Observatory, AlbaNova, 106 91 Stockholm, Sweden ({\tt fredrik@astro.su.se})}
\altaffiltext{5}{Division of Geological and Planetary Sciences, Mail Code 150-21, California Institute of Technology, Pasadena, CA 91125 ({\tt gab@gps.caltech.edu})}
\altaffiltext{6}{Division of Physics, Mathematics and Astronomy, Mail Code 150-24, California Institute of Technology, Pasadena, CA 91125 ({\tt acab@astro.caltech.edu})}
\altaffiltext{7}{Max Planck Institut f\"{u}r Astronomie, Koenigstuhl 17, 69117 Heidelberg, Germany ({\tt dullemon@mpia.de})}
\altaffiltext{8}{Department of Astronomy, University of Texas at Austin, 1 University Station, C1400, Austin, TX 78712 ({\tt nje, jes@as.utexas.edu})}

\begin{abstract}
Mid-infrared (23--35~$\mu$m) emission from the deeply embedded ``Class
0'' protostar IRAS~16293-2422 is detected with the Spitzer Space
Telescope infrared spectrograph. A detailed radiative transfer model
reproducing the full spectral energy distribution (SED) from 23~$\mu$m
to 1.3~mm requires a large inner cavity of radius 600~AU in the
envelope to avoid quenching the emission from the central
sources. This is consistent with a previous suggestion based on high
angular resolution millimeter interferometric data. An alternative
interpretation using a 2D model of the envelope with an outflow cavity
can reproduce the SED but not the interferometer visibilities. The
cavity size is comparable to the centrifugal radius of the envelope
and therefore appears to be a natural consequence of the rotation of
the protostellar core, which has also caused the fragmentation leading
to the central protostellar binary. With a large cavity such as
required by the data, the average temperature at a given radius does
not increase above 60--80~K and although hot spots with higher
temperatures may be present close to each protostar, these constitute
a small fraction of the material in the inner envelope. The proposed
cavity will also have consequences for the interpretation of molecular
line data, especially of complex species probing high temperatures in
the inner regions of the envelope.
\end{abstract}


\section{Introduction}
In the earliest stages of their evolution, low-mass young stellar
objects are thought to be deeply embedded in an envelope of cold gas
and dust. The objects therefore emit large fractions of their emission
at far-infrared through submillimeter wavelengths.  \cite{andre93}
introduced the term ``Class 0'' for such protostars with $L_{\rm
  submm}/L_{\rm bol} > 0.5\%$ where $L_{\rm submm}$ is the luminosity
originating longwards of 350~$\mu$m. The typical mid-infrared
10--20~$\mu$m fluxes of these sources are predicted to be very weak,
below 1~mJy. With the Spitzer Space Telescope's high sensitivity, it
now becomes possible to search for these objects in the mid-infrared
and test models for their structure. This paper presents Spitzer/IRS
observations of the deeply embedded low-mass protostar
IRAS~16293-2422. These observations, together with a detailed dust
radiative transfer model \citep[][]{schoeier02,hotcorepaper},
constrain the properties of the innermost region of the protostellar
envelope.

In recent years, a substantial step forward in studies of deeply
embedded protostars is the quantitative characterization of the
physical structure of their envelopes through observations with
submillimeter cameras combined with dust radiative transfer models
\citep[e.g.,][]{shirley00,shirley02,motte01,jorgensen02}. The
submillimeter observations are predominantly sensitive to the cold
material in the outer envelope, which is why the radiative transfer
models constrain the structure of the envelope, such as its density
and temperature profiles, on larger scales. Thus, the mid-infrared
data probing the inner warm region are highly complementary.

Of all the protostars studied in this manner, IRAS~16293-2422 has
gained particular attention: it is a protobinary with a separation of
$\sim$~800~AU \citep{mundy92} and a total luminosity of
27~$L_\odot$. With this luminosity, material in the inner $\sim$100~AU
of the envelope will have temperatures higher than $\sim$~100~K due to
the heating from the central protostar(s)
\citep{schoeier02}. Interpretations of, e.g., the chemistry in these
regions rely on extrapolation of the larger scale envelope parameters
to small scales. A way to test the extrapolation is through
aperture-synthesis observations of the dust continuum emission
compared to the predictions from the models. \cite{hotcorepaper}
analyzed 1~mm observations from the Owens Valley Radio Observatory
Millimeter Array and suggested that the envelope around
IRAS~16293-2422 in fact has an inner cavity with a size corresponding
to the well resolved binary. The projected binary separation of 800~AU
would suggest a minimum inner radius for the envelope of 400~AU
because the binary would clear a cavity through dynamical
interaction. Still, these data are mostly sensitive to the cold dust
in the envelope. Independent constraints can be obtained at
mid-infrared wavelengths where it is possible to study the warm dust
close to the central protostar. IRAS~16293-2422 is also known to drive
prominent outflows \citep{walker88}, and the associated cavities may
further allow the mid-infrared radiation to escape.

\section{Observations}\label{obs}
IRAS~16293-2422A was observed by the ``Cores to Disks'' (c2d) legacy
team \citep{evans03} using the infrared spectrograph (IRS) on 2004
August 29 (\dataset[ads/sa.spitzer\#11826944]{AOR key 11826944}).  The
source was observed in staring mode with the SH and LH modules
($9.9-19.6\ \mu$m and $18.7-37.2\ \mu$m; $R \sim 600$) and SL1 and SL2
modules ($5.2-14.5\ \mu$m; $R \sim 64-128$). The observations were
centered on the south eastern component in the protostellar binary at
$\alpha_{2000}=16^{\rm h}32^{\rm m}22\fs87$,
$\delta_{2000}=-24^\circ28'36\farcs1$ \citep{looney00}.

Data reduction started from the BCD images using S10.5.0 Spitzer
archive data. The processing includes bad-pixel correction,
extraction, defringing and order matching using the c2d analysis
pipeline (Lahuis et al. in prep.). The source is not detected
shortward of 23~$\mu$m but rises steeply from 23--35~$\mu$m
(Fig.~\ref{onedcavity}). For these wavelengths an optimized source
profile extraction was used to separate the source from the extended
cloud emission. H$_2$ and PAH emission is also detected: both are
extended and not directly associated with the source.

\section{Dust radiative transfer model}\label{results}
\subsection{Background}\label{1dstruc}
\begin{figure}
\resizebox{0.75\hsize}{!}{\plotone{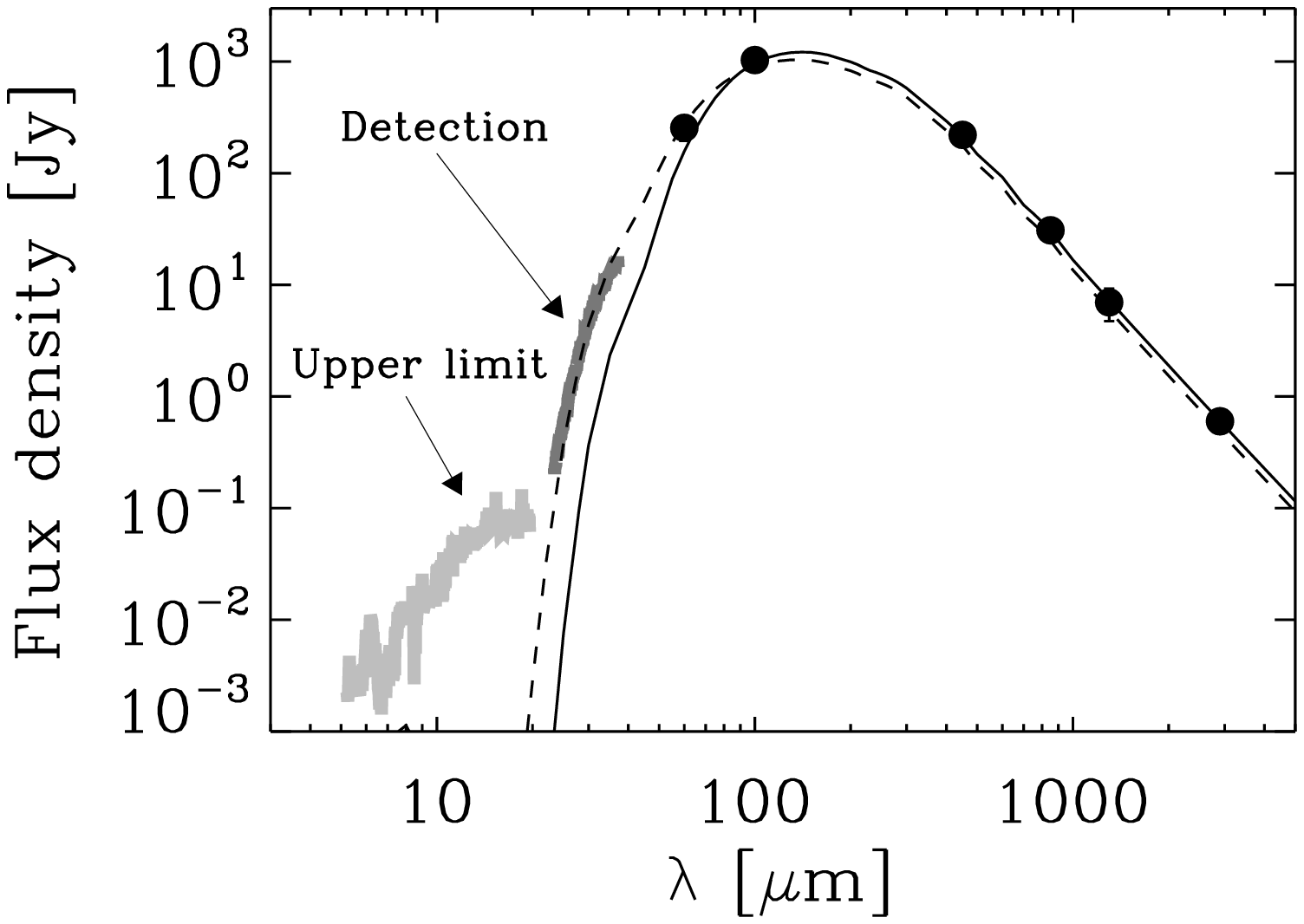}}
\resizebox{0.75\hsize}{!}{\plotone{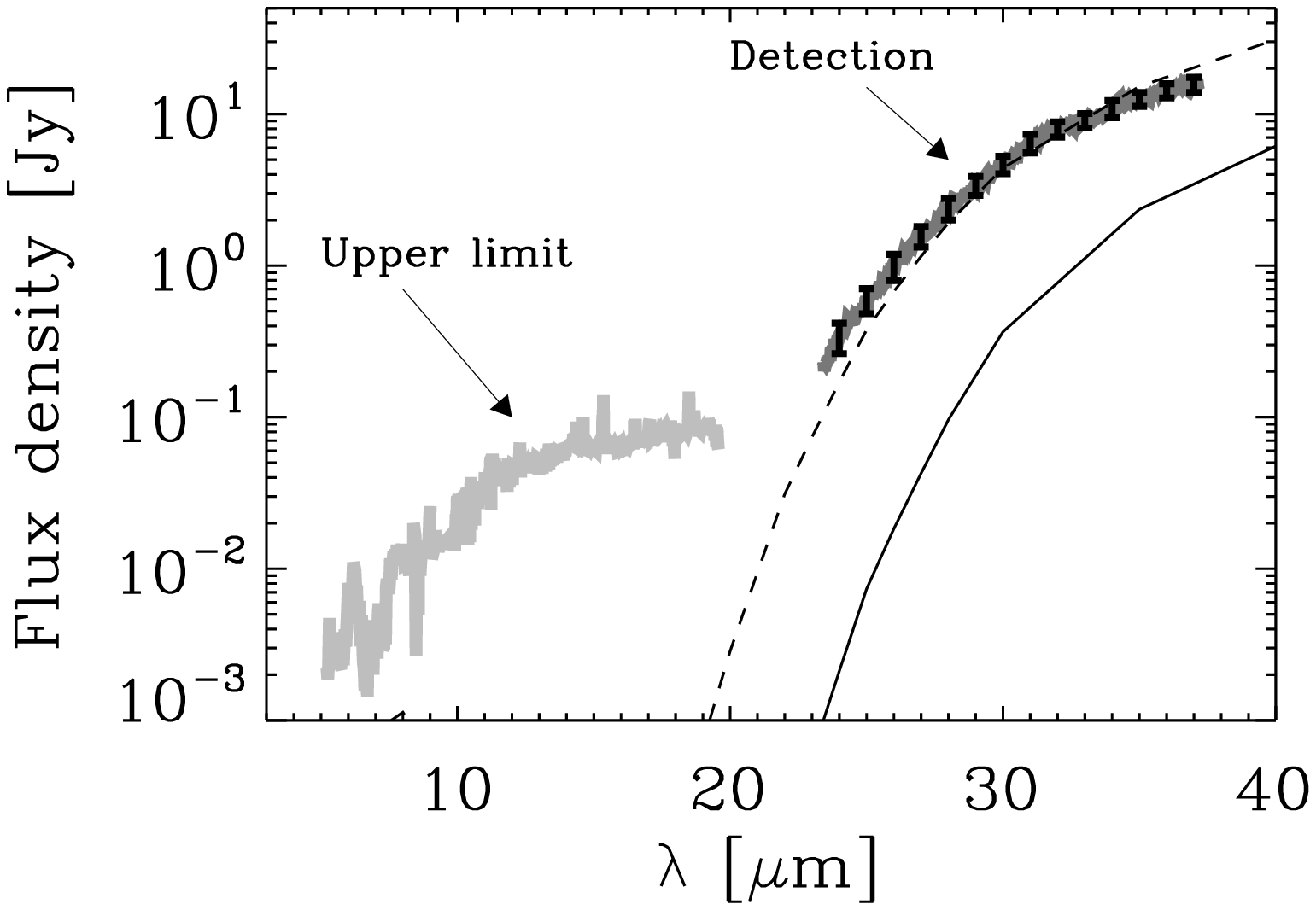}}
\caption{Spitzer/IRS observations of IRAS~16293-2422 and models for
  its SED. In both panels the dark grey line indicates the IRS
  detection and the light grey line the upper limit. The black solid
  line shows the model of \cite{schoeier02}, the dashed line the best
  fit model from this paper. In the upper panel the symbols indicate
  the far-infrared/submillimeter SED from \citeauthor{schoeier02}. In
  the lower (blow-up) panel the symbols indicate the bins of the IRS
  spectrum used in the fits.}\label{onedcavity}
\end{figure}

The model for the outer envelope structure by \cite{schoeier02} cannot
reproduce the mid-infrared flux seen by Spitzer
(Fig.~\ref{onedcavity}) and we therefore reexamine these models. In
short, such models calculate the dust radiative transfer through a
spherical envelope illuminated by a central source supplying the total
luminosity $L$ using the 1D radiative transfer code \emph{Dusty}
\citep{dusty}. The inputs into the models are the spectrum of the
central source (approximated as a blackbody with a specific effective
temperature, $T_{\rm eff}$), the inner and outer radii of the
envelope, $R_{\rm i}$ and $R_{\rm e}$, and the density distribution of
the envelope. The density distribution is typically described by a
power-law profile, $\rho = \rho_0\, (r/r_0)^{-p}$, leaving two free
parameters: the slope $p$ and normalization, $\rho_0$. The latter can
be specified through the total dust optical depth at a specified
wavelength (e.g., $\tau_{100}$ for the optical depth at
100~$\mu$m). The spectrum of the central source and the inner radius
of the envelope are not constrained by the submillimeter emission. As
in \cite{schoeier02} and \cite{jorgensen02} we adopt the opacities of
coagulated dust grains with thin ice mantles at a density of
$10^6$~cm$^{-3}$ from \cite{ossenkopf94}. For further discussions of
the models we refer to the cited papers.

The detection of IRAS~16293-2422 in the mid-infrared suggests that the
envelope surrounding the source is less optically thick at these
wavelengths compared to the model of \cite{schoeier02}, which severely
underestimates the emission at mid-infrared wavelengths
(Fig.~\ref{onedcavity}). Its dust optical depth toward the central
illuminating source is $\approx 40$ at 25~$\mu$m. One option to
reproduce the entire SED is to increase the inner radius of the
envelope. This is consistent with the high angular resolution
millimeter interferometric data by \cite{hotcorepaper} who suggested
the presence of an inner cavity in the envelope with a radius of
$\approx$~400~AU resolved by the interferometer. With such a cavity,
the column density of material along the line of sight is decreased,
whereas the total mass of the envelope and thereby the submillimeter
emission remains largely unaffected.

\subsection{Results}
To test the idea of a larger inner radius, a grid of models is
calculated varying $R_{\rm e}$ and $\tau_{100}$ as in
\citeauthor{schoeier02} - but now also letting $R_{\rm i}$ vary. The
slope of the density distribution, $p$, is fixed to 1.9, which
provides a good fit to the brightness distribution in single-dish
observations on a few thousand AU scales and aperture synthesis
observations on a few hundred AU scales
\citep{schoeier02,hotcorepaper}. To perform the fits and estimate the
$\chi^2$ statistic, the mid-infrared spectrum is rebinned into
fourteen $\Delta \lambda=1~\mu$m bins. The error in each measurement
is taken to be the standard deviation of the flux measurements in each
bin with an addition of 10\% of the resulting flux, representing the
uncertainty in the absolute flux calibration and relative calibration
across the spectrum.

\begin{figure}
\plotone{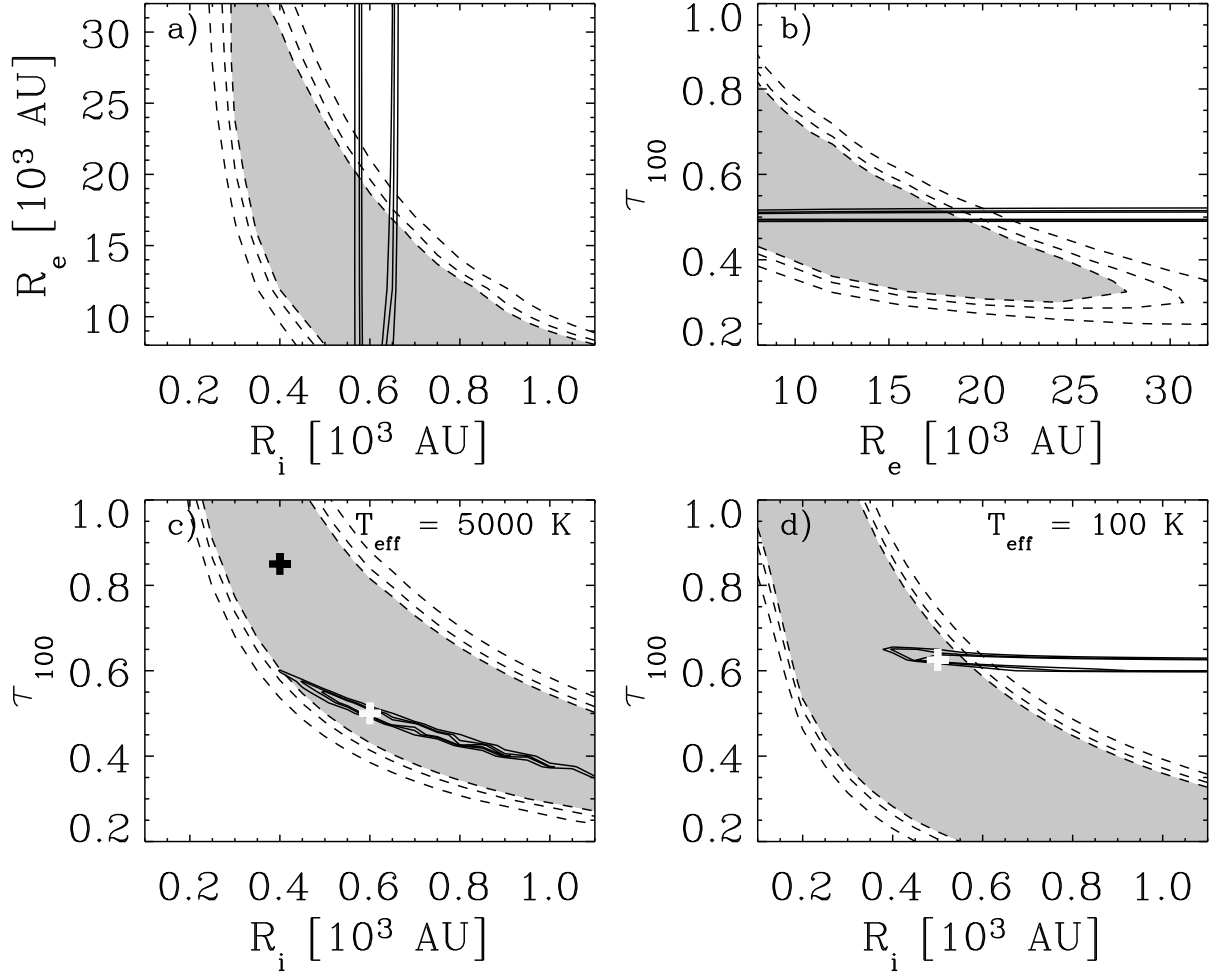}
\caption{$\chi^2$-confidence plots from fits to the SED with the
  envelope model. In each panel the contours indicate the 90\%, 95\%
  and 99\% confidence limits. The dashed line contours indicate the
  fits to the far-infrared/submillimeter SED \citep{schoeier02}
  whereas the solid line contours indicate the fits to the Spitzer/IRS
  mid-infrared SED. The grey colored region indicate the 90\%
  confidence region for the fits to the submillimeter SED. In panel
  a)-c) the effective temperature of the central blackbody is 5000~K,
  in d) it is taken to 100~K. The best fit to the combined
  submillimeter and mid-infrared datasets is shown in panel c) and d)
  with the white ``+''. In c) the model of \cite{hotcorepaper} fitting
  the submillimeter SED and interferometer data is shown with the
  black ``+''.}\label{paramplot}
\end{figure}

The $\chi^2$ confidence plots are shown in Fig.~\ref{paramplot}. As
panels a)-c) illustrate, the mid-infrared SED provides good
constraints on the size of the inner cavity whereas it is completely
insensitive to the emission from the outermost regions of the
envelope. Both the submillimeter and mid-infrared SEDs are well fitted
by a model similar to that presented in \cite{schoeier02} with the
additional constraint from the IRS spectrum that $R_{\rm i} =
600_{-150}^{+250}$~AU (corresponding to
$\tau_{100}=0.5_{-0.10}^{+0.05}$) (Fig.~\ref{paramplot}c). The reduced
$\chi^2$ for the best fit model is $\approx 2.5$ and $\lesssim 1$ for
the fits to the mid-infrared and far-infrared/submillimeter parts of
the SED, respectively. A cavity of 400~AU radius suggested by the
model of \cite{hotcorepaper} is not ruled out by the mid-infrared
observations (Fig.~\ref{paramplot}). That model fits the submillimeter
SED and millimeter interferometer data which resolve the envelope
cavity. Still, the mid-infrared data place stronger constraints on the
thickness of the envelope in that particular case compared to just
considering the submillimeter SED.

One natural concern is whether the spectral shape of the central
source influences the result. In particular, the central circumstellar
disks could be bright at mid-infrared wavelengths compared to the
stellar blackbodies. Still, if the envelope is highly optically thick
at 23--35~$\mu$m the effect of changing the SED of the central source
will be small. Fig.~\ref{paramplot}d illustrates the extreme example
where the total luminosity of the central object is coming from a
100~K blackbody. The change in spectrum affects the larger scale
emission of the envelope somewhat in the optically thin (long
wavelength) regime but the requirement that the envelope only is
marginally optically thick at 23--35~$\mu$m still constrains the
minimum cavity size.

\subsection{Going to 2D?}\label{2dstruc}
\begin{figure}
\plotone{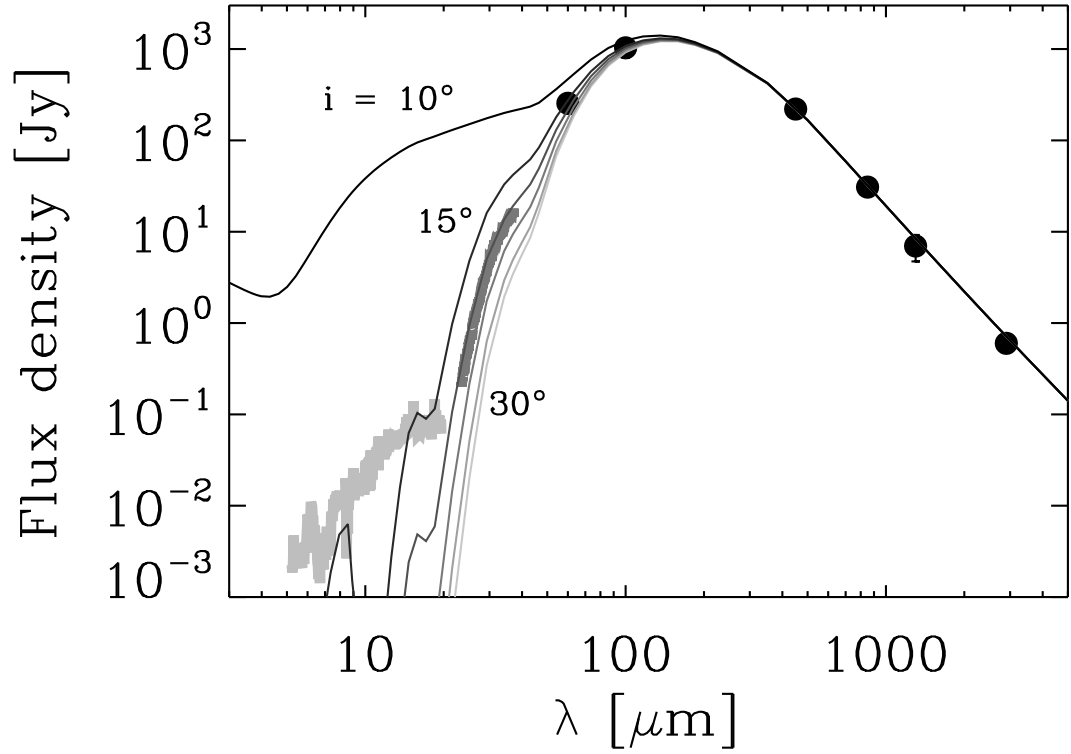}
\caption{Envelope models with $R_{\rm i} = 32$~AU and a conical
  outflow cavity with an opening angle of 30$^\circ$ and inclination
  angles of 10$^\circ$, 15$^\circ$, 17$^\circ$, 20$^\circ$, 25$^\circ$
  and 30$^\circ$ (going from dark to light grey). The IRS detection
  and upper limits are shown by the dark and light grey lines as in
  Fig.~\ref{onedcavity}.}\label{2dcomp}
\end{figure}

Another question is whether the 1D models truly represent the
structure of the envelopes at the level of detail suggested by these
observations. An alternative explanation for the increased
mid-infrared radiation is that the 1D models are not a good
representation of the structure of the envelope, since they do not
contain any outflow cavities. \cite{whitney03a} showed that the
mid-infrared SED strongly depends on the inclination angle if a large
outflow cavity is present in the envelope. To illustrate the
sensitivity to the inclination angle, a 2D radiative transfer model
was calculated using a dust Monte Carlo code
\citep{dullemond00,dullemond04}. In this model, a conical slice of the
envelope is cut out mimicking an outflow cavity. The cavity is assumed
to have an opening angle of 30$^\circ$ and to be completely devoid of
material. Otherwise the 2D structure is identical to the 1D model of
\cite{schoeier02}, i.e., extending inwards to about
30~AU. Fig.~\ref{2dcomp} compares the SED for the 2D model with
varying inclination angles. For lines-of-sight down or close to the
cavity the mid-infrared flux is indeed increased and a good fit to the
mid-infrared SED is obtained with an inclination angle of
17--20$^\circ$.

\begin{figure}
\plotone{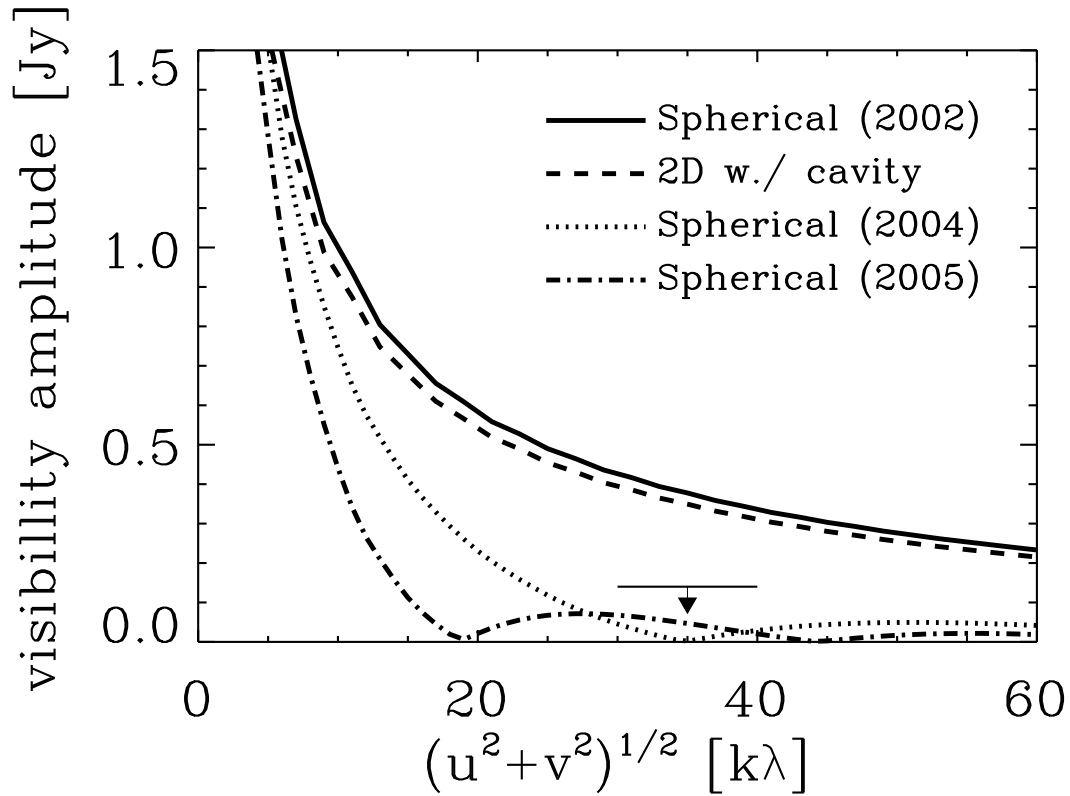}
\caption{Visibility amplitudes versus projected baseline lengths at
  1.37~mm. The phase center is taken to be the center point of the
  circumbinary envelope, i.e., between the two binary components
  \citep[see][]{hotcorepaper}. Shown are the 1D spherical model of
  \cite{schoeier02} (solid line), the models of \cite{hotcorepaper}
  and this paper with larger inner cavities (dotted and dashed-dotted,
  respectively) and the 2D outflow cavity model (dashed). Note that
  the outflow cavity and spherical models with $R_{\rm i} = 32$~AU are
  almost indistinguishable and overproduce the emission at the
  30--40~k$\lambda$ baselines compared to the OVRO observations of
  \cite{hotcorepaper} (indicated by the horizontal line and downward
  arrow).}\label{onedcavityuv}
\end{figure}

Introducing the outflow cavity, however, does not alleviate the need
for an inner cavity inferred from the high angular resolution
millimeter data: Fig.~\ref{onedcavityuv} compares the predicted
visibilities at 1.37~mm from the 1D envelope models with increasing
cavity sizes and the 2D model presented above. The 1D spherical model
with a small inner cavity and the similar model with an outflow cavity
are virtually indistinguishable and both produce too much flux on
intermediate baselines (10--30~k$\lambda$). In contrast the
visibilities from the models of \cite{hotcorepaper} and this paper
drop quickly at these projected baselines, as required by comparison
to the interferometer data.

The mid-infrared and high angular resolution millimeter data are
complementary: the best constraints on the 1D cavity model are
obtained from the mid-infrared data probing the warm dust but the
existence of the inner cavity is demonstrated by the millimeter
interferometer data (which also provides constraints on the
circumstellar disks embedded in the envelope). It should be emphasized
that the data do not exclude the presence of an outflow cavity, but
that an inner cavity is still required even though an outflow cavity
can reproduce the mid-infrared data. An even more complicated
multi-dimensional structure can also not be excluded.

\section{Discussion}\label{discuss}
The inner cavity in the IRAS~16293-2422 envelope inferred here is of
comparable size to the separation of the binaries. Does this simply
reflect the centrifugal radius due to rotation in the core before the
collapse into the binary? \cite{zhou95} examined large scale CS
single-dish observations and found that two regimes of rotation
applied to the IRAS~16293-2422 core indicative of differential
rotation: the inner region of the envelope rotating with $\Omega =
3\times 10^{-13}$~s$^{-1}$ and the outer region about factor 6
slower. A similar rotation rate in the inner envelope was found
through single-dish observations by \cite{narayanan98}. From
comparison of the line observations to the \cite{terebey84} models of
slowly rotating, infalling cores, \citeauthor{zhou95} found a
centrifugal radius (where gravity is balanced by rotation) of 600~AU
(corresponding to a collapse timescale of 1$\times 10^5$~years in the
\cite{terebey84} model for the sound speed derived by
\citeauthor{zhou95}). This similarity to the inner radius derived from
the mid-infrared observations is striking: the requirement of an inner
envelope cavity from the strength of the mid-infrared emission may
simply reflect the centrifugal radius of the protostellar envelope.

\cite{boss99} examined the properties of rotating collapsing cores. In
these simulations cores with similar ratios between their rotation and
gravitational energy such as IRAS~16293-2422 ($\beta=0.03$; based on
the estimates of the rotation rate by \cite{zhou95} and the envelope
structure in this paper) do indeed collapse and form a binary. It
therefore seems that the conclusion about the inner cavity in the
IRAS~16293-2422 envelope is a natural consequence of its rotation,
which has also resulted in the protostellar binary. On the other hand,
if the binary had formed by other means, the dynamical interaction
between the binary components and the envelope would likely have
resulted in clearing of the inner cavity.

The temperatures and densities in the outer envelope are very similar
in both models with small and large inner cavities. The interpretation
of the chemistry of molecular species probing these regions will
therefore not be affected by the revised models in \cite{hotcorepaper}
and this paper. The interpretation of line emission coming from the
inner ``hot core'' \citep[e.g.,][]{ceccarelli00a,schoeier02} will
clearly be affected, however. In the cavity model preferred here, the
envelope temperature does not rise above 60--80~K. Locally, of course,
the temperature will be higher since the binary separation is
comparable to the cavity size. In particular, at the edge of the
cavity close to each of the binary components the temperature should
be higher than that obtained from the spherically symmetric
calculations, whereas in orthogonal directions to the binary the
temperature is expected to be lower. However, since each binary
component can heat the envelope material above 100~K only within about
100~AU, it will be a small fraction of the envelope which has
temperatures higher than 100~K typical of a ``hot core''. It therefore
appears likely that the complex organic species have a different
origin than passively heated material in the protostellar envelope. An
alternative could be accretion shocks in each of the circumstellar
disks or, as suggested by the high angular resolution submillimeter
observations by \cite{chandler05}, a result of the shock associated
with the IRAS~16293-2422A outflow.

\acknowledgments The research of JKJ was supported by NASA Origins
Grant NAG5-13050. Support for this work, part of the Spitzer Legacy
Science Program, was also provided by NASA through contract 1224608
issued by the Jet Propulsion Laboratory, California Institute of
Technology, under NASA contract 1407. Astrochemistry research in
Leiden is supported by a NWO Spinoza grant and a NOVA grant. FLS
acknowledges financial support from the Swedish Research Council.

\clearpage

\clearpage

\clearpage

\clearpage

\end{document}